# Enhancement of Scattering Efficiency and Development of Optical Magnetometer Using Quantum Measurement Set Up

Sufi O Raja and Anjan K Dasgupta*

Department of Biochemistry University of Calcutta, 35 Ballygunge Circular Road, Kolkata-700019, India

**ABSTRACT**: Quantum measurement principle is employed to detect water quality and presence of nano-colloids. The setup uses spatially low coherent light source, for which the outcome of measurement is dependent on the presence of a reflecting surface and a linear polarizer. The introduction of a reflecting surface induces enhanced side scattering. The enhancement has specific patterns for pure water, ions and nanoparticles and can be employed to detect refractive index of liquids at high sensitivity. The differential enhancement can be used as an optical magnetometer that sensitively senses magnetic moments of colloidal magnetic nanoparticles at concentration untenable by other measurement techniques.

Quantum measurement (QM) deals with collapse of wave function[1]. While most of the published works on QM deal primarily with the formal challenges, we use an accepted undebated norm of the QM, namely the intricate dependence of the measured entity with the process of measurement. In the process we probe chemical optical and magnetic properties of the local environment. Our basic inspiration is the double slit experiment. By introducing a reflecting surface an enhancement of the scattered light is observed if and only if a polarizer is introduced in the optical path[2] despite the fact the polarizer cuts off the light intensity.

We have used a linearly polarized light source, which has low spatial coherence. This type of source is used in Low spatially coherent Enhanced Back Scattering (LEBS) technique [3,4], where back scattering efficiency increases due to constructive interference between two time reversed path pairs. LEBS is found in polydispersed media. Here we have studied the effect of a reflecting surface that led to altered side scattering efficiency of several liquid samples and report the attenuation of enhancement upon introduction of polydispersity (nanoparticles) and ions. The non-classical phenomenon that was noted is the effect of polarizer in the light path. The reflector could enhance the side scattering efficiency when the polarizer is present, and the enhancement effect practically disappears in absence of the polarizer.

The used reflecting surface (R) was a phosphor bronze plate. The distance between R and sample holder was 0.3 cm. The plane (XZ) of reflection was perpendicular to the direction of light propagation (XY). Schematic illustration of the used experimental setup is shown in Figure 1.

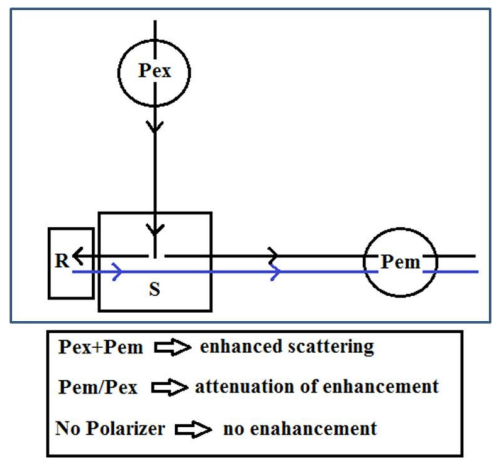

**Figure 1**: Schematic representation of the used experimental setup: Where Pex is excitation polarizer and Pem is emission polarizer. R is reflecting surface and S is sample holder containing liquids. The arrows indicate the direction of light propagation. Blue line represents the reflected light beam.

At 470 nm the illumination power of Xenon-arc lamp is highest and the PMT detector response is also maximum for that wavelength. Hence, we chose that wavelength to study the R induced differential scattering. We performed time scan for 120 seconds at 470 nm in absence and presence of R. For analysis purpose we took the ratio of intensity at 470 nm in presence (+R) and absence (-R) of R and we assigned the ratio as enhancement factor (F).

F=intensity at 470 nm (+R)/intensity at 470 nm (-R)

We used four different excitation and emission polarizer angle combinations that are 90/90, 90/0, 0/0 and 0/90 (angles expressed in °) along with three different combinations that are using no polarizers, only excitation polarizer and only emission polarizer. The F values for above mentioned combinations are provided in Table 1. We found maximum value of F (R induced maximum enhancement of scattering) for 90/90 excitation/emission polarizer angle combination and lower value of F upon use of only one polarizer and no enhancement in absence of polarizers.

To check whether the reflection is solely responsible for the observed effect or not, we took different reflecting surfaces.

Use of white paper as reflecting surface increased the scattering efficiency than phosphor bronze and use of aluminum foil as reflecting surface further increased the scattering efficiency. When we chose a paper with surface rubbed to increase roughness, we found attenuated response. Figure 2 shows the effect of different R on scattering efficiency.

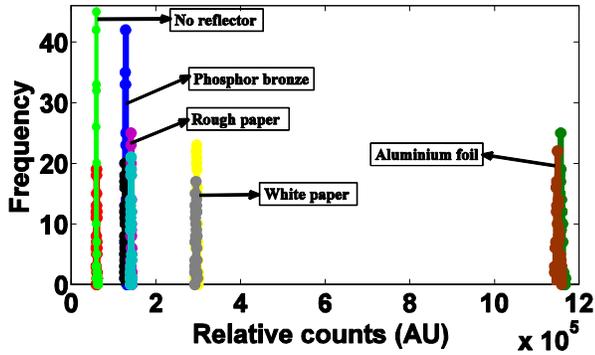

**Figure 2:** Enhanced scattering efficiency of de-ionized water in presence of Reflector (R). The histograms indicate the temporal fluctuation of scattering at 470 nm. Time scan was performed at 470 nm for 120 sec in presence of both polarizers in absence (red and green are duplicate set) and in presence (the other colors specific to R in duplicates) of 'R'.

Next, we studied the effect of presence of ion and nanoparticles on R induced enhanced side scattering efficiency. Figure 3 shows the intensity histograms for de-ionized water and water containing same concentration of ionic iron in ionic and nanoform (iron oxide nanoparticles). For 10 µM ferrous chloride solution and colloidal iron oxide, we get lower value of F than de-ionized water (see Figure 3).

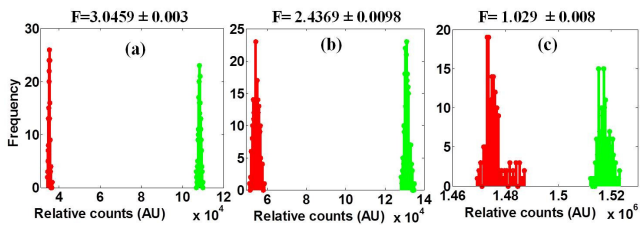

**Figure 2:** Enhanced scattering efficiency of de-ionized water in presence of Reflector (R). The histograms indicate the temporal fluctuation of scattering at 470 nm. Time scan was performed at 470 nm for 120 sec in presence of both polarizers in absence (red and green are duplicate set) and in presence (the other colors specific to R in duplicates) of 'R'.

The results discussed so far suggested that the observed enhancement of scattering efficiency is caused by constructive interference between scattered light, reflected light and reflected light induced scattered light beams. The observed differential scattering efficiency in presence of R is due to differential contribution of coherent and incoherent phase relations of the contributing beams in the path between exit points from sample holder and detector. The extent of spatial coherence is directly related to slit width of excitation and emission slits (see Figure 4a) due to variation of coherent length (LSC). The dependence of enhancement factor on slit width also indicates that the observed enhanced scattering efficiency is due to constructive interference between primary side scattered beam and secondary forward scattered (induced by reflected light) beam. In presence of R the medium is exposed to dual light source and results in double scattering. Mechanistically, the observed phenomenon is similar to LEBS, where constructive interference between two time reversed path pairs (double scattering) is the reason of observed enhanced back scattering[3]. But in LEBS, polydispersed medium is required to enhance the back scattering. That means in LEBS, the medium itself plays the role of a reflector. Here we used a reflecting surface to create the double scattering (the R acts as a secondary light source) induced enhanced side scattering, where optically transparent medium (de-ionized liquid water) yields maximum extent of enhancement. Introduction of ions or nanoparticles introduces incoherency in the phase relation between the contributing beams, which is the reason behind the observed lower value of F than de-ionized water. Basically the technique measure optical transparency of the studied medium.

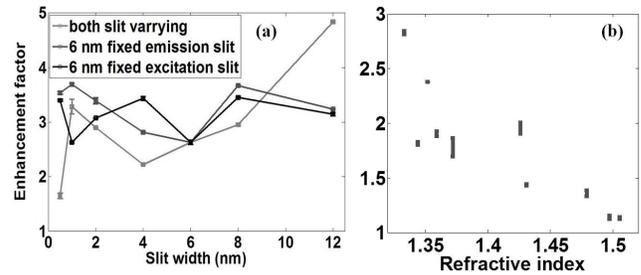

**Figure 4:** (a) Dependence of enhancement factor on excitation/emission slit width, where both slits were varied (pale gray points), excitation slit varied keeping emission slit fixed at 6 nm (dark gray points) and emission slit was varied keeping excitation slit fixed at 6 nm (black points). Here the R was of phosphor bronze and distance between R and sample holder was 0.3 cm. (b) Dependence of enhancement factor on liquid properties: Variation of enhancement factor (for 470 nm) with solvent refractive index.

In open quantum system the interaction with the environment (environment induced de-coherence) is the reason of collapse of quantum superposition (coherence)[5]. Here we are measuring the extent of de-coherence induced by the local environment (the solvent) on the spatial coherent relation of reflected and scattered light beam. Here, the de-coherence is introduced by the local medium itself and extent of de-coherence is reflected in differential extent of enhancement. Higher optical transparency means lower velocity retardation of the light beams (lower value of refractive index of the medium), which results in higher value of F due to lower extent of incoherent phase relation between the contributing beams.

Now, the non-classicality of the work can be realized from absence of enhancement in absence of polarizers. The absence of enhancement when polarizers are not present can be explained assuming that the unpolarized scattered light is partially polarized due to reflection [6,7]. From the principles of the quantum eraser double slit experiment [8-10] we know that interference will be destroyed if one of the interfering wave fronts pass through a polarizer. The argument evidently does not occur if polarizer is present before each of the interfering beams. The enhancement can be accounted for by this line of argument (see Table 1). Incidentally, the extent of increase of F upon introduction of smoother surface (see Figure 2) also follows similar argument as in this case the percentage of reflected light available for constructive interference is increased. In this context we should mention that the extent of enhancement of scattering efficiency is also depends on surface area of the sample holder as we found lower value of F upon use of 1ml cuvette (data not shown). This also implies that the F solely depends on the amount of reflected light.

If the R induced observed enhanced scattering efficiency is due to constructive interference then the response should be sensitive to any change in the Refractive Index (RI). For that purpose we used nine non-aqueous solvents (acetone, acetonitrile, di-methyl formamide, di-methyl sulfoxide, ethyl acetate, ethyl ether, n-Hexane, o-Xylene and toluene) having different RI values and performed polarized scattering experiments in absence and in presence of R. The RI of the used solvents is given in supporting information (Table S1). Figure 4 (b) shows that the extent of enhanced scattering efficiency decreases with solvent RI (Figure 4b).Change in RI and polarity can change the electrical dipole-electrical dipole coupling and results in differential scattering efficiency [11]. Increase in RI means retardation of velocity, which changes the phase relation between scattered and reflected light [12,13], which further results in spatial incoherency due to contribution of destructive interference. In this context, we can say the observed lowering of F value in presence of ions is due to increase in RI of the water, as presence of solute increases the RI of water (R).

Next, we studied the effect of temperature variations. De-ionized liquid water is a special solvent due to presence of dynamic structural domains [14-16] resulted from van der Waals interaction and hydrogen bonding network. We performed temperature dependency of said response for de-ionized water. The direct evidence of role of structured water on scattering enhancement comes from thermal dependency of F. Figure 3d shows that the enhancement factor is maximum in the temperature range of 25-36C. Temperature induced increase in F may be associated with the decrease in RI with temperature [17]. However, decrease of enhancement factor beyond 36C cannot be explained by RI. According to literature increase of temperature increases the monomeric water population and self ionisation [18]. And we found that the enhancement factor decreases in presence of ions (see Figure 3b). Hence, we may conclude that the observed reduction in enhancement factor is due to decrease in polarity resulting from temperature induced self ionisation.

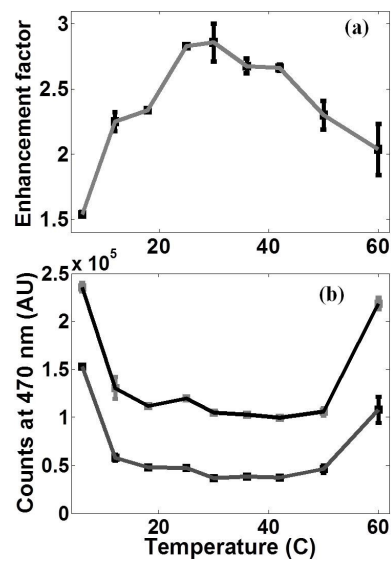

**Figure 5:** (a) Temperature dependency of R induced enhanced scattering efficiency of milli-Q water at 90°/90° excitation/emission polarizer angle combination. (b) Relative counts at 470 nm in absence (gray line) and in presence (black line) of R.

At low temperature the structured domains of de-ionized water causes higher RI due to higher velocity retardation. But with increasing temperature the oligomeric domains breaks and the liquid water become slightly transparent to light and results in reduced RI. The observed temperature optimum is observed due to higher optical transparency of water in that temperature region. Figure 5 (b) shows that the relative scattering intensity is lowest in that region. This implies that in that temperature region percent of reflected light is higher and that is why we found higher value of F. The reciprocal pattern of Figure 5 (a) and (b) clearly indicates that higher optical transparency of the medium (lower value of scattering intensity in absence of R) results in higher value of R induced enhanced scattering efficiency (F).

Based on our observation that we can measure optical transparency of a medium, we tried to develop an optical magnetometer to measure the magnetic moment of magnetic nanoparticles. We incubated the magnetic colloids on a 0.5 Tesla magnet for one hour. The logic behind this magnetic pulling experiment is that magnetic nanoparticle with higher magnetic moment will be pulled down more by the magnet and the solution will be more transparent to light. We used three different iron oxide nanoparticles with different surface modification that are aminopropyl tri-ethoxy silane (APTES) coated, citrate coated and dextran coated iron oxide nanoparticles. The detail of synthesis protocol and hydrodynamic size distribution of the synthesized nanoparticles is provided in the supplementary section. From SQUID analysis we found that the order of magnetic moment is citrate coated > APTES coated > dextran coated (see Figure 6 a-c).

As we did not observe any significant extent of enhancement for 10 μM (ionic iron) citrate coated iron oxide nanoparticles solution, we used nM concentration to observe R in-duced differential scattering efficiency for the used three iron oxide nanoparticles. Figure 6 (d-f) shows the intensity histograms in absence and presence of R.

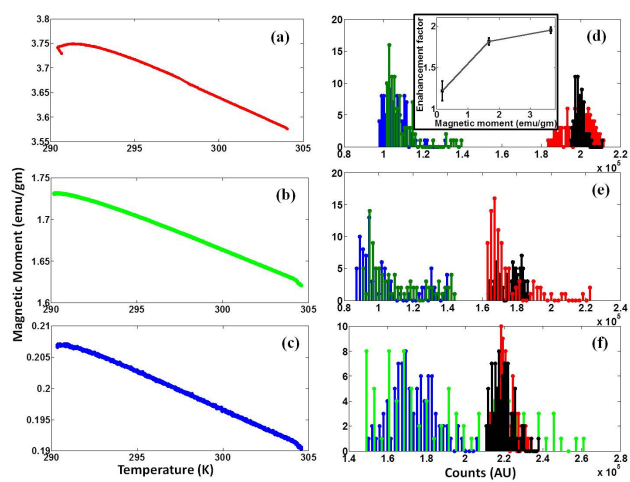

**Figure 6:** Left panel shows the magnetic moment vs temperature plots for citrate (a), APTES (b) and dextran (c) coated iron oxide nanoparticles. The right panel shows the histograms of R induced enhanced scattering intensity for 470 nm (at 90°/90° ex/em polarizer angle). The blue and green colors indicate duplicate sets of intensity histogrmas in absence of R and red and black clours indicate duplicate sets of inteensity histograms in presence of R for citrate (e), APTES (b) and dextran (f) coated iron oxide nanoparticles. The inset on the panel shows the variation of F with magnetic moment of the three nanoparticles shown in panels (a)(b) and(c).

Intensity histograms (6 d-f) clearly show that R induced enhancement of scattering efficiency after magnetic pulling is highest for the particle, which has highest magnetic moment. The inset in the Figure 6 (d) shows the relation between magnetic moment of magnetic nanoparticles and R induced enhancement of scattering (F). The only parameter which can result in non-linear relation between moment and F is the presence of ions as presence of trace amounts of ions can reduce the enhancement factor (see Figure 3b). The ions in solution may come from free surface coating agent and leached iron ions due to improper surface coating. Hence, a membrane filtration step is required to remove the free ions and to increase the sensitivity level of the proposed optical technique to measure magnetic moment. As the citrate coated nanoparticle is more prone to iron leaching, the presence of trace iron ions (leached during sample preparation and incubation) results in the observed attenuated response than expected. In addition the sensitivity of the technique can be increased by using smoother reflecting surface.

The proposed approach, in its application side is thus useful in determining the RI of liquids, purity of water. The magnetic moment of magnetic nanoparticles can also be measured by the above said technique.

In conclusion, we have employed the basic principle of quantum measurement problem in exploring liquid state properties. The virtual absence enhancement of scattering in absence of both polarizers implies that the observation is an implicit function of the measurement process, a typical signature of the quantum measurement problem. In each of the polarizer and reflector combination the detector unravels the state to a new configuration. The outcome of the measurement is obviously different in each case. The paper highlights the implicit relation between measurement induced collapse of states and the chemical information extractable as a result. The sensitivity of R induced enhancement of scattering to trace ions seems to be of special interest, as this indirectly implies restructuring of hydrogen bond network in presence of trace ions. Finally the observations suggest some unique structure and polarity features of water at 25°-37° C, a temperature range supportive of several biological functions. In addition the method enables detection of colloidal nanomaterials at concentration that a several order magnitude lower than what is permissible using dynamical light scattering. Again, an optical magnetometer to measure magnetic moment of magnetic nanoparticles can be developed using this quantum measurement set up.

**Table 1.** Scattering enhancement factor (F) using various excitation and emission polarizer angle combinations.

| Pex | Pem | F |
| --- | --- | --- |
| Absent | Absent | 1.1639±0.0174 |
| Absent | 90° | 2.2351±0.0244 |
| 90° | Absent | 1.3977±0.0099 |
| 90° | 90° | 2.8286±0.0144 |
| 90° | 0° | 1.7053±0.0966 |
| 0° | 0° | 1.3929±0.0584 |
| 0° | 90° | 1.8384±0.0606 |

Pex: excitation polarizer, Pem: emission polarizer, F: enhancement factor.

## ASSOCIATED CONTENT

### Supporting Information

The supporting information contains the details of materials and methods, RI values of the used solvents, synthesis and hydrodynamic size distribution of magnetic nanoparticles.

## AUTHOR INFORMATION

### Corresponding Author

Prof. Anjan K Dasgupta

### Present Addresses

Department of Biochemistry, University of Calcutta, 35 Ballygunge Circular Road, Kolkata-700019, India. E-mail: adgcal@gmail.com

Author Contributions

‡These authors contributed equally.

Notes

The authors declare no competing financial interests.

REFERENCES

(1) Zurek, W. H. *Reviews of Modern Physics* **2003**, *75*, 715.

(2) Falkenburg, B. In *EPSA Philosophical Issues in the Sciences*; Springer: 2010, p 31.

(3) Kim, Y. L.; Liu, Y.; Wali, R. K.; Roy, H. K.; Backman, V. *Applied optics* **2005**, *44*, 366.

(4) Radosevich, A. J.; Mutyal, N. N.; Turzhitsky, V.; Rogers, J. D.; Yi, J.; Taflove, A.; Backman, V. *Optics letters* **2011**, *36*, 4737.

(5) Zurek, W. H. *arXiv preprint quant-ph/0306072* **2003**.

(6) Chen, H.-S.; Rao, C. R. N. *Journal of Physics D: Applied Physics* **1968**, *1*, 1191.

(7) Jones, B. F.; Fairney, P. *Image and Vision Computing* **1989**, *7*, 253.

(8) Holden, M.; McKeon, D.; Sherry, T. *Canadian Journal of Physics* **2011**, *89*, 1079.

(9) Scully, M. O.; Englert, B.-G.; Walther, H. *Nature* **1991**, *351*, 111.

(10) Walborn, S.; Cunha, M. T.; Pádua, S.; Monken, C. *Physical Review A* **2002**, *65*, 033818.

(11) Brahma, S.; Howard Jr, W.; Nelson, W. *The Journal of Physical Chemistry* **1984**, *88*, 5551.

(12) Bernardo, L. M.; Soares, O. D. D. *Journal of Modern Optics* **1988**, *35*, 1857.

(13) Zwerdling, S. *JOSA* **1970**, *60*, 787.

(14) Miyazaki, M.; Fujii, A.; Ebata, T.; Mikami, N. *Science* **2004**, *304*, 1134.

(15) Shin, J.-W.; Hammer, N.; Diken, E.; Johnson, M.; Walters, R.; Jaeger, T.; Duncan, M.; Christie, R.; Jordan, K. *Science* **2004**, *304*, 1137.

(16) Smith, J. D.; Cappa, C. D.; Wilson, K. R.; Messer, B. M.; Cohen, R. C.; Saykally, R. J. *Science* **2004**, *306*, 851.

(17) Bashkatov, A. N.; Genina, E. A. In *Saratov Fall Meeting 2002: Optical Technologies in Biophysics and Medicine IV*; International Society for Optics and Photonics, p 393.

(18) Bandura, A. V.; Lvov, S. N. *Journal of physical and chemical reference data* **2006**, *35*, 15.

Supplementary Informations

# Enhancement of Scattering Efficiency and Development of Optical Magnetometer Using Quantum Measurement Set Up

Sufi O Raja and Anjan K Dasgupta*

Materials and Methods:

*Sample preparation:*

All de-ionized water samples were filtered through a 0.2 micron negatively charged membrane filter prior to spectroscopic measurement.

*Scattering measurement:*

Light scattering studies were performed in a PTI fluorimeter (QuantamasterTM40) with a 72 W Xenon lamp. We used software controlled synchronously varying excitation and emission monochromators and Glan Thompson polarizer for polarized light scattering studies. The temperature inside the sample holder was controlled by software controlled Peltier.

For unpolarized light we fixed the slit width at 2 nm (excitation) and at 4 nm (emission) and for polarized light we used slit widths of 6 nm (excitation and emission) for all spectroscopic measurement except slit width variation experiment.

*Magnetic measurements and magnetic pulling experiment:*

Magnetic moment of the synthesized nanoparticles was measured using SQUID magnetometer (Quantum design, PPMS). To measure the magnetic moment at 298 K, field cooling (FC) experiment at 100 Oe field from 290 to 305 K was performed.

80 μL 3.75 μM particle solutions were placed on a magnet (0.5 Tesla) for 1 hour at 25 °C. Then 10 μL aliquots from each eppendorf were taken and mixed with 2 ml de-ionized filtered water. Finally the polarized scattering measurements were performed in absence and presence of reflector.

**Table S1: The used solvents and their physical parameters:**

| Solvents | Polarity Index | Refractive Index | Source and purity |
|---|---|---|---|
| Acetone | 5.1 | 1.359 | SRL (99.5%) |
| Acetonitrile | 5.8 | 1.344 | SRL (99.8%) |
| Dimethyl formamide | 6.4 | 1.431 | SRL (99%) |
| Dimethyl sulfoxide | 7.2 | 1.479 | Merck (99.9%) |
| Ethyl acetate | 4.4 | 1.372 | Spectrochem (99.8%) |
| Ethyl ether | 2.8 | 1.352 | SRL (99.5%) |
| n-Hexane | 0.1 | 1.426 | Merck (96%) |
| o-Xylene | 2.5 | 1.505 | SRL (98%) |
| Toluene | 2.4 | 1.497 | SRL (99.8%) |
| Water | 10.2 | 1.333 | Millipore |

**Sources:**

Polarity: http://macro.lsu.edu/howto/solvents/Polarity%20index.htm

Refractive index: http://www.stenutz.eu/chem/solv23.php

Synthesis of colloidal magnetic nanoparticles:

*Synthesis of citrate coated iron oxide nanoparticles:*

Colloidal Iron oxide NPs were synthesized through co-precipitation of ferrous and ferric salts. Briefly, Ferric chloride ($FeCl_3$, $6H_2O$) and ferrous chloride ($FeCl_2$, $4H_2O$) are taken in 2:1(w/w) ratio and dissolved in 2 N HCl (2 ml solution containing 1gm $FeCl_3$ and 0.5 gm $FeCl_2$) and co-precipitated by one step addition of 1.5 M sodium hydroxide (NaOH) solution (40 ml). Then the black precipitate stirred for half an hour at room temperature (25° C). All the steps were performed in inert nitrogen environment. The formed precipitate was washed several times (6-7 times) by distilled water through magnetic decantation. Citrate functionalized Iron oxide NPs were synthesized by addition of tri-sodium citrate and citric acid mixture (1:2 w/w ratios, 5 ml) in washed precipitate (3 ml) and 4 hours of moderate stirring at 25° C. Then pH was adjusted to ~7.2 by addition of 1.5 M NaOH solution. Finally suspensions were centrifuged at 10,000 rpm for 12 minutes to remove the larger aggregates.

*Synthesis of APTES coated nanoparticles:*

The bare iron oxide nanoparticles were synthesized using the protocol described in Section 2.2.1. After vigorous washing, 3 ml of bare nanoparticles solution was taken and mixed with 1 ml diluted (25 μL 12 N HCL + 1 ml milli-Q water) HCl solution and stir for 15 mins to break the large clusters. Then freshly prepared 2 ml water-APTES mixture (1:1 v/v) was added. The mixture was stirred for 3 hours at room temperature and for 30 mins at 60 ºC.

*Synthesis of Dextran coated nanoparticles:*

Briefly, 1.5 gm dextran and 230 mg $FeCl_3$ was dissolved in 2 ml de-ionized water. 153 mg $FeCl_2$ was dissolved in 1 ml de-ionized water and store at 4 ºC for 10 mins. Then $FeCl_2$ solution was added to the $FeCl_3$-polymer mixture with a syringe at stirring condition in nitrogen envi-ronment. After addition cold 8 ml liquor ammonia was added and stirred for 2 hours at 85 °C to remove the excess ammonia.

*Dynamic Light Scattering (DLS) study:*

Hydrodynamic size and zeta potential of synthesized Iron oxide NPs were measured using DLS (Nano ZS, Malvern, equipped with a 4 mW He-Ne Laser ($\lambda$= 632 nm)). Briefly, 10 µL of synthesized nanoparticles was diluted to 1 ml by Milli-Q water and filtered through 0.2 micron membrane filter prior to measurement.

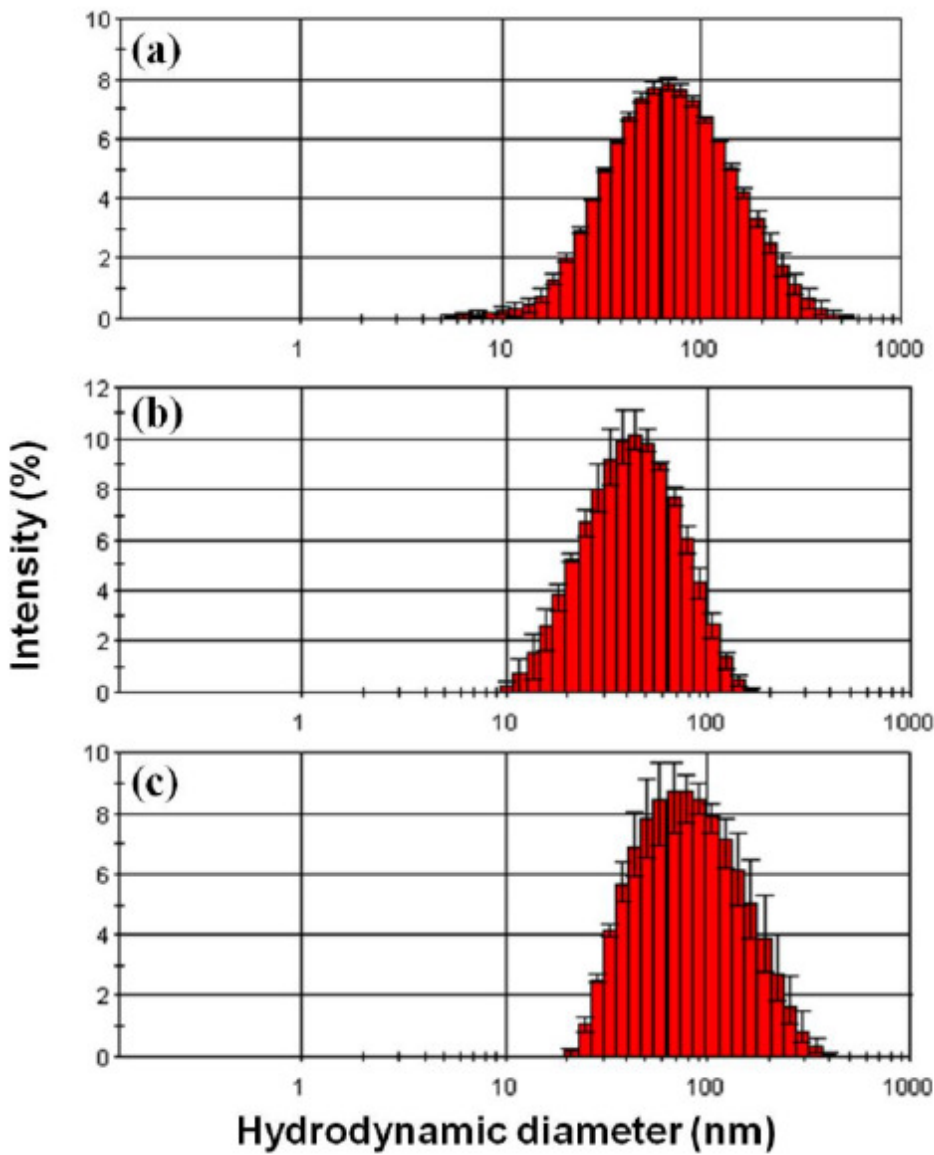

Figure S1: Hydrodynamic size distribution (% intensity) of the synthesized iron oxide nanoparticles, (a) aminopropyl-triethoxy silane (APTES) coated, (b) citrate coated and (c) Dextran coated nanoparticles.